\documentclass[11pt,twoside,a4paper]{article}
\usepackage[bookmarksnumbered, colorlinks, plainpages]{hyperref}
\usepackage{amsthm}
\usepackage{latexsym}
\usepackage{amsmath}
\usepackage{epsfig}
\usepackage{times}
\usepackage{amssymb}

\newtheorem{thr}{\quad Theorem}
\newtheorem{lem}{\quad Lemma}

\newtheorem{defin}{\quad Definition}

\newtheorem{exam}{\quad Example}

\title{Existence and Solution-representation of IVP for LFDE with generalized Riemann-Liouville fractional derivatives and $n$ terms}

\author{Myong-Ha Kim, Guk-Chol Ri and Hyong-Chol O\\ \\
\small\textit{Department of Mathematics,} \\
\small\textit{Kim Il Sung University,  Pyongyang D.P.R.Korea}}

\date{First version submitted Feb 12, 2013; last verion revised Nov 3, 2013\\ to appear in FCAA Vol.17 No.1 2014}

\begin{document}
\maketitle
%\hline

\begin{abstract}
This paper provides the existence and representation of solution to an initial value problem for the general multi-term linear fractional differential equation with generalized Riemann-Liouville fractional derivatives and constant coefficients by using operational calculus of Mikusinski's type. We prove that the initial value problem has the solution of if and only if some initial values should be zero.



\end{abstract}

%\hline
%
%-----------------------1. Introduction-----------------
%
\section{Introduction}
In recent years there have been considerable interests in the theory and applications of fractional differential equations (\cite{CDFP, hil1, KO, MCVXF, SAT, sha}). Indeed, it is well known that fractional derivative is excellent tool for description of memory and heredity effects \cite{MKM}. It is therefore important to understand the advantages of using fractional derivative in classical equations. 

Applications of fractional calculus require fractional derivatives of different kinds \cite{hil1}. There exist several different definitions of fractional differentiation \cite{KST, pod}.
The (right-sided) \textit{Riemann-Liouville fractional derivative} is defined by
\begin{equation}
D^\alpha_{0+}f(x):=\frac{d^n}{dx^n}I^{n-\alpha}_{0+}f(x)~,~x>0,    \label{1}
\end{equation}
where\\
\begin{equation}
I^\alpha_{0+}f(x):=\frac{1}{\Gamma(\alpha)}\int_0^x(x-t)^{\alpha-1}f(t)dt~,~x>0,    \label{2}
\end{equation}
\begin{equation*}
I^0_{0+}f(x):=f(x)~,~x>0
\end{equation*}
is the (right-sided) \textit{Riemann-Liouville fractional integral} of order $\alpha$ with lower limit 0 (see e.g. \cite{KST, pod}).
The (right-sided) \textit{Caputo fractional derivative} is defined by
\begin{equation}
^cD^\alpha_{0+}f(x):=I^{n-\alpha}_{0+}\frac{d^n}{dx^n}f(x)~,~x>0    \label{3}
\end{equation}
whenever the right-hand side exists (see e.g. \cite{KST, pod}).

In \cite{hil2} the \textit{generalized Riemann-Liouville fractional derivative} (GRLFD) of order $\alpha$ and type $\beta$  is defined as
\begin{equation}
D^{\alpha, \beta}_{0+}f(x):=(I^{\beta(n-\alpha)}_{0+}\frac{d^n}{dx^n}(I^{(1-\beta)(n-\alpha)}_{0+}f))(x)~,~x>0,    \label{4}
\end{equation}
where $n-1 < \alpha \leq n \in \mathbf{N}$ and $0 \leq \beta \leq 1$, whenever the right-hand side exists(see e.g. \cite{HLT}). This generalization gives the classical Riemann-Liouville fractional differential operator (1) if $\beta =0$. For $\beta=1$ it gives the Caputo fractional differential operator (3). The concept of GRLFD is appeared in the theoretical modeling of broadband dielectric relaxation spectroscopy for glasses \cite{hil3}. Some properties and applications of the GRLFD are given in \cite{hil1, hil2, SMT, ST, THS, TSMD}. Several authors (see \cite{FKT, SMT, TSMD}) called (4) the \textit{Hilfer fractional derivative} or \textit{composite fractional derivative} operator. 

More recently, new results for the fractional differential equations with GRLFDs were obtained in \cite{FKT, SMT, TSMD, tom}. In \cite{tom} the authors provided an approach based on the equivalence of a nonlinear Cauchy type problem with GRLFD to a nonlinear Volterra integral equation of the second kind  in spaces of summable functions on a finite interval of the real axis and proved uniqueness and existence of the solution using a variant of the Banach's fixed point theorem. In \cite{TSMD} they investigated the solution of space–time fractional diffusion equations with a generalized Riemann–Liouville time fractional derivative.

The operational calculus was successfully applied in ordinary differential equations, integral equations, partial differential equations and in the theory of special functions (see \cite{luc, YL}). For example, in \cite{LS} Mikusinski's operational calculus has been applied to solve a Cauchy boundary-value problem for a certain linear equation involving the Riemann-Liouville fractional derivatives. In \cite{LG} by using the operational calculus of Mikusinski's type for the Caputo fractional differential operator they obtained exact solutions of an initial value problem for linear fractional differential equations with constant coefficients and Caputo fractional derivatives.

In \cite{HLT} an operational calculus of the Mikusinski's type was introduced for the generalized Riemann-Liouville fractional derivative operator and it was applied to solve the corresponding initial value problem for the general $n$-term linear fractional differential equation with constant coefficients and  GRLFD of arbitrary orders and types. In \cite{HLT} they provided initial values dependent on the type of generalized Riemann-Liouville fractional derivatives in $n$-term linear fractional differential equation (see the problem (36) and (37) of \cite{HLT}) and defined the space of solutions dependent on the type(see the definition 1 of \cite{HLT}). They provided a result of existence and uniqueness of the solution of n-terms linear fractional differential equation with the orders of the fractional derivatives are all between $n-1$ and $n$ and they gave the formula of the solution (see the formula (42) of \cite{HLT}). 

Let consider the following initial value problem
\begin{equation*}
D^{0.7, 0}_{0+}y(x)-D^{0.5,0}_{0+}y(x)=0,~I^{0.3}_{0+}y(0)=C_{0} \in \mathbf{R},~I^{0.5}_{0+}y(0)=C_{1} \in \mathbf{R}.
\end{equation*}
This initial value problem DOES not belong to the class of the problems that were solved in \cite{HLT} at all because this problem has 2 terms and the orders of the derivatives in this equation are between 0 and 1 and not between 1 and 2.

The aim of this paper is to EXTEND the results obtained in \cite{HLT} to a {\it wider class of the equations including the above mentioned equation} by using the method of operational calculus developed in \cite{HLT}. We present the existence and representation of solution for the $n$-term linear initial value problem with generalized Riemann-Liouville fractional derivatives and constant coefficients. According to our result, the solution to the above mentioned problem exists if and only if the initial value $C_{1}$ is zero.

We start with some preliminaries and operational calculus for the generalized fractional derivative in Section 2 mainly according to \cite{HLT}. In Section 3, we provide a necessary and sufficient condition for existence of the solution of the initial value problem and some examples.
%%-2. Preliminaries and Operational calculus for generalized fractional derivatives--------------------------
%
\section{Preliminaries and Operational calculus for Generalized fractional derivatives}
\cite{luc, YL} contain the background and basic knowledge of the resultsof this paper.
In this section, all materials are from \cite{HLT} unless specially indicated.

%-----------------Definition 1--------------------------
\begin{defin}\cite{HLT}
 A function $f(x),~x>0$, is said to be in the space $C^{m}_{\alpha},~m \in \mathbf{N}_{0}=\mathbf{N} \cup \{0\}, \alpha\in\mathbb{R}$, if there exists a real number $p,~p>\alpha$, such that
\begin{equation}
f^{(m)}(x)=x^{p}f_{1}(x)                         \label{5}
\end{equation}
with a function $f_{1}(x)$ in $C[0,~\infty)$. Especially the space $C^{0}_{\alpha}$ is denoted by $C_{\alpha}$.
\end{defin}
By \cite{HLT}, $C_{\alpha}$ is a vector space and the set of spaces $C_{\alpha}$ is ordered by inclusion according to 
\begin{equation}
C_{\beta} \subset C_{\alpha} \Leftrightarrow \alpha \geq \beta.    \label{6}
\end{equation}

%-----------------Lemma 1--------------------------
%
\begin{lem}\cite{LG}
The Riemann-Liouville fractional integral $I^{\alpha}_{0+},~\alpha \geq 0$, is a linear map of the space $C_{\gamma},~\gamma \geq -1$, in to itself , that is,
\begin{equation}
I^{\alpha}_{0+}: C_{\gamma} \to C_{\alpha + \gamma} \subset C_{\gamma}.   \label{7}
\end{equation}
\end{lem}
It is well known, that the operator $I^{\alpha}_{0+},~\alpha \geq 0$ has the following convolutional representation in the space $C_{\gamma},~\gamma \geq -1$.
\begin{equation}
I^{\alpha}_{0+}f(x)=h_{\alpha}(x) \star f(x) =\int_{0}^{x}f(x)h_{\alpha}(x-t)dt~,~x>0,    \label{8}
\end{equation}
\begin{equation}
h_{\alpha}(x) :=\frac{x^{\alpha -1}}{\Gamma(\alpha)}~,~x>0,~\alpha>0,   \label{9}
\end{equation}
where 
\begin{equation}
(g \star f)(x) :=\int_{0}^{x}g(x-t)f(t)dt~,~x>0,     \label{10}
\end{equation}
is the Laplace convolution. Moreover, the following properties of the Riemann-Liouville fractional integral are well known:
\begin{equation*}
I^{\alpha}_{0+}I^{\beta}_{0+}f(x)=I^{\beta}_{0+}I^{\alpha}_{0+}f(x)=I^{\alpha+\beta}_{0+},~\alpha,~\beta \geq 0,
\end{equation*}
\begin{equation}
\underbrace{(I^{\alpha}_{0+} \cdots I^{\beta}_{0+}f)(x)}_n=(I^{n \alpha}_{0+}f)(x).     \label{11}
\end{equation}
%-----------------Lemma 2--------------------------
%
\begin{lem}\cite{LG}
Let $f \in C^{m}_{-1},~m \in \mathbf{N}_{0},~f(0)= \cdots =f^{(m-1)}(0)=0$ and $g \in C^{1}_{-1}$. Then the Laplace convolution
\begin{equation}
h(x) :=\int_{0}^{x}f(t)g(x-t)dt                \label{12}
\end{equation}
is in the space $C^{m}_{-1}$ and $h(0)= \cdots =h^{(m)}(0)=0$.
\end{lem}

Now we define the following space of functions according to \cite{HLT} and use in operational calculus.
%-----------------Definition 2--------------------------
%
\begin{defin}\cite{HLT}
A function $y \in C_{-1}$ is said to be in the space $\Omega ^{\alpha}_{\beta}$, if $I^{(1-\beta)(n-\alpha)}_{0+}y \in C^{n}_{-1},~n-1<\alpha \leq n \in \mathbf{N}$ for $\alpha>0,~0 \leq \beta \leq 1$.
\end{defin}
Then the following lemmas hold.
%-----------------Lemma 3--------------------------
%
\begin{lem}\cite{HLT}
Let $y \in \Omega ^{\alpha}_{\beta},~n-1<\alpha \leq n \in \mathbf{N}$. Then the Riemann-Liouville fractional integral (2) and the generalized fractional derivative (4) are connected by the relation
\begin{equation}
(I^{\alpha}_{0+}D^{\alpha,\beta}_{0+}y)(x)=y(x)-y_{\alpha,\beta}(x),~x>0,   \label{13}
\end{equation}
where
\begin{equation}
y_{\alpha,\beta}(x):=\sum _{k=0}^{n-1} \frac{x^{k-n+\alpha-\beta \alpha +\beta n}}{\Gamma(k-n+\alpha-\beta \alpha +\beta n+1)} \frac{d^{k}}{dx^{k}}(I^{(1-\beta)(n-\alpha)}_{0+}y)(0+),~x>0.   \label{14}
\end{equation}
\end{lem}

As in the case of Mikusinski's type operational calculus for the Riemann-Liouville or for the Caputo fractional derivatives we have the following lemmas (see e.g. \cite{HLT, LS, LG})
%-----------------Lemma 4--------------------------
%
\begin{lem}\cite{HLT}
The space $C_{-1}$ with the operations of the Laplace convolution and ordinary addition becomes a commutative ring $(C_{-1},~\star,~+)$ without divisors of zero.
\end{lem}

This ring can be extended to the field $M_{-1}$ of convolution quotients by following the lines of the classical Mikusinski's operational calculus;
\begin{equation}
M_{-1}:=(C_{-1} \times (C_{-1} \setminus \{0\})) / \sim,   \label{15}
\end{equation}
where the relation "$\sim$" is defined as usual by
\begin{equation*}
(f,~g) \sim (f_{1},~g_{1}) \Leftrightarrow (f \star g_{1})(x)=(g \star f_{1})(x).
\end{equation*}
 This relation is an equivalence relation in terms of Lemma 4. 
For the sake of convenience, the elements of the field $M_{-1}$ can be formally considered as convolution quotients $\frac{f}{g}$.
The operations of addition and multiplication are the defined in $M_{-1}$ as usual:
\begin{equation}
\frac{f}{g} + \frac{f_{1}}{g_{1}}:= \frac{f \star g_{1}+g \star f_{1}}{g \star g_{1}}   \label{16}
\end{equation}
and
\begin{equation}
\frac{f}{g} \cdot \frac{f_{1}}{g_{1}}:= \frac{f \star f_{1}}{g \star g_{1}}. \label{17}
\end{equation}

Then the space $M_{-1}$ with the operations of addition (16) and multiplication (17) becomes a commutative field $(M_{-1},~\cdot,~+)$, where the unit element is defined by $I:=\frac{f}{f}$,and the zero element is defined by $0:=\frac{0}{f},~f \neq 0$ \cite{HLT}.
The flied $(M_{-1},~\cdot,~+)$ is the quotient flied of the ring $(C_{-1},~\star,~+)$.
The ring $(C_{-1}$ can be embedded into the field $M_{-1}$ by the map:
\begin{equation*}
f \mapsto (f,~I)= \frac{h_{\alpha} \star f}{h_{\alpha}},~\alpha>0.
\end{equation*}
\begin{equation}
s_{\alpha}:= \frac{I}{h_{\alpha}}=\frac{h_{\alpha}}{h_{\alpha} \star h_{\alpha}}=\frac{h_{\alpha}}{h_{2\alpha}},~\alpha>0. \label{18}
\end{equation}
%%----------------Lemma 5--------------------------
%%
\begin{lem}\cite{HLT}
For the function $y \in C_{-1}$, the Riemann-Liouville fractional integral of $y$ can be represented multiplication in the field $M_{-1}$ :
\begin{equation}
(I^{\alpha}_{0+}y)(x)=\frac{I}{s_{\alpha}}  \cdot y.     \label{19}
\end{equation}
And the generalized fractional derivative $D^{\alpha,\beta}_{0+}y$ of the function $y \in \Omega ^{\alpha}_{\beta},~\alpha>0,~0 \leq \beta \leq 1$, can be represented as multiplication in the field $M_{-1}$ of convolution quotients
\begin{equation}
(D^{\alpha,\beta}_{0+}y)(x)=s_{\alpha} \cdot y - s_{\alpha} \cdot y_{\alpha,\beta}. \label{20}
\end{equation}
\end{lem}

Formula (11) means that for $\alpha >0,~n \in \mathbf{N}$
\begin{equation*}
h^{n}_{\alpha}(x):=\underbrace{h_{\alpha} \star \cdots \star h_{\alpha}}_n=h_{n \alpha}(x).
\end{equation*}
This relation can be extended to an arbitrary positive real power exponent:
\begin{equation*}
h^{\lambda}_{\alpha}(x):=h_{\lambda\alpha}(x).
\end{equation*}
For any $\lambda >0$, then inclusion $h^{\lambda}_{\alpha}(x) \in C_{-1}$ hold true and the following relations can be easily proved
\begin{equation}
h^{\beta}_{\alpha} \star h^{\gamma}_{\alpha}=h_{\beta \alpha} \star h_{\gamma \alpha}=h_{(\beta+\gamma)\alpha}= h^{\beta+\gamma}_{\alpha},~h^{\beta}_{\alpha_1}=h^{\gamma}_{\alpha_2} \Leftrightarrow \alpha_1\beta=\alpha_2\gamma.         \label{21}
\end{equation}
The above relations motivate the following of the element $s_{\alpha}$ with an arbitrary real power exponent $\lambda$.
\begin{equation}
s^{\lambda}_{\alpha}:=\left\{
\begin{array}{rl}
h^{-\lambda}_{\alpha} & \text{if} \quad\lambda<0,\\
I & \text{if}  \quad\lambda=0,\\
\frac{I}{h^{-\lambda}_{\alpha}} & \text{if}  \quad\lambda>0.
\end{array} \right.   \label{22}
\end{equation}
For any $\alpha,~\alpha_1,~\alpha_2,~\beta,~\gamma>0$, it follows from this definition and the relations (21) that 
\begin{equation*}
s^{\beta}_{\alpha} \cdot s^{\gamma}_{\alpha}=s_{(\beta+\gamma)\alpha}= s^{\beta+\gamma}_{\alpha},~s^{\beta}_{\alpha_1}=s^{\gamma}_{\alpha_2} \Leftrightarrow \alpha_1\beta=\alpha_2\gamma.
\end{equation*}
 For the application of the operational calculus to solution of differential equations with generalized fractional derivatives it is important to identify the elements of the field $M_{-1}$, which can be represented by the elements of the ring $C_{-1}$.
One useful class of such representations is given by the following lemma (see e.g. \cite{HLT,  LG}).
%
%-----------------Lemma 6-------------------------
\begin{lem}\cite{HLT, LG}
Let $\beta>0,~\alpha_{i}>0, i=1,\cdots ,n$. Then 
\begin{equation}
\frac{s^{\beta}_{\alpha}}{I-\sum^{n}_{i=1}}=x^{\beta\alpha_1}E_{(\alpha_1\alpha, \cdots \alpha_n\alpha), \beta\alpha}(\lambda_1 x^{\alpha_1\alpha}, \cdots, \lambda_n x^{\alpha_n\alpha})                \label{23}
\end{equation}
with the multivariate vector-index Mittag-Leffler function 
\begin{equation*}
E_{(a_1, \cdots, a_n), b}(z_1, \cdots, z_n):=\sum_{k=0}^{\infty} \sum_{l_1+ \cdots +l_n=k\\ l_1,\cdots, l_n \geq 0}(k, l_1,\cdots,l_n)\frac{\prod_{j=1}^n z^{l_j}_j}{\Gamma(b+\sum^{n}_{j=1}a_j l_j)}, 
\end{equation*}
\begin{equation}
\newline ~a_j>0,~b>0,~z_j \in \mathbf{C}    \label{24}
\end{equation}
and multinomial coefficients
\begin{equation*}
(k, l_1,\cdots,l_n):=\frac{k!}{l_1! \times \cdots \times l_n!}.
\end{equation*}
\end{lem}

When $a \in \mathbf{Z}$ and $b \in \mathbf{Z}$ satisfy $a \leq b$, we denote by $\mathbf{Z}_{a,b}$ the set of all integers $i$ satisfying $a \leq i \leq b$.
%%%%%%-----------------3.Existence and uniqueness of the solution-------------------------

\section{Existence and uniqueness of the solution}
In this section the constructed operational calculus is applied to solve linear fractional differential equations with generalized derivatives and constant coefficients.
We consider the linear differential equation
\begin{equation}
D^{\alpha_0,\beta_0}_{0+}y(x)-\sum^n_{i=1}a_i D^{\alpha_i,\beta_i}_{0+}y(x)=g(x)  \label{25}
\end{equation}
and the initial conditions
\begin{equation}
\frac{d^k}{dx^k}I^{(1-\beta_i)(m_i-\alpha_i)}_{0+}y(0+)=y_{k-(1-\beta_i)(m_i-\alpha_i)} \in \mathbf{R},~ i \in \mathbf{Z}_{0,n},~k \in \mathbf{Z}_{0,m_i-1},  \label{26}
\end{equation}
where $\alpha_i \geq 0, m_i-1<\alpha_i \leq m_i \in \mathbf{N},~0 \leq \beta_i \leq 1,~ i \in \mathbf{Z}_{0,n}$ and $a_i \in \mathbf{R},~i \in \mathbf{Z}_{1,n}$.
The ordering 
\begin{equation}
m_0-(1-\beta_0)(m_0-\alpha_0) \geq \cdots \geq m_n-(1-\beta_n)(m_n-\alpha_n)  \label{27}
\end{equation}
is assumed without lost  of generality.
%%%%%----------------------Definition3----------------------------------
\begin{defin}
The function $\bigcap ^n_{i=0} \Omega ^{\alpha_i}_{\beta_i}$ is called the solution of the initial value problem (25) and (26), if $y$ satisfied the equation (25) and the initial value (26).
\end{defin}
%-----------Theorem 1-----------------
\begin{thr}
The initial value problem (25) and (26) with homogeneous initial conditions
\begin{equation}
\frac{d^k}{dx^k}I^{(1-\beta_i)(m_i-\alpha_i)}_{0+}y(0+)=0,~ i \in \mathbf{Z}_{0,n},~k \in \mathbf{Z}_{0,m_i-1},     \label{28}
\end{equation}
where 
\[g(x) \in \left\{
\begin{array}{rl}
C_{-1} & \text{if}\quad \alpha_0 \in \mathbf{N}, \\
C^1_{-1} & \text{if}\quad \alpha_0 \notin \mathbf{N} \\
\end{array}	 \right.\]
has a solution $y(x)$ which is unique in the space $\bigcap ^n_{i=0} \Omega ^{\alpha_i}_{\beta_i}$ and provided by
\begin{equation}
y(x)=\int^{x}_{0}t^{\alpha_0-1}E_{(\alpha_0-\alpha_1, \cdots, \alpha_0-\alpha_n), \alpha_0}(a_1 t^{\alpha_0-\alpha_1}, \cdots, a_n t^{\alpha_0-\alpha_n})g(x-t)dt.  \label{29}
\end{equation}
\end{thr}  
%----------------Proof
%%begin{pr}
~\textbf{Proof:} ~Let $y(x) \in \bigcap ^n_{i=0} \Omega ^{\alpha_i}_{\beta_i}$ and satisfied (25) and (28). Then the following algebraic equation in the field $M_{-1}$ of convolution quotients is obtained
\begin{equation}
s_{\alpha_0} \cdot y-\sum_{i=1}^n a_i s_{\alpha_i} \cdot y=g    \label{30}
\end{equation}
from homogeneous initial value problem (25) and(28) by using the Lemma 5.
This linear equation has a unique solution in the field $M_{-1}$
\begin{equation}
y=\frac{I}{s_{\alpha_0} -\sum_{i=1}^n a_i s_{\alpha_i}} \cdot g.    \label{31}
\end{equation}
From (23) we get 
\begin{align}
\nonumber\frac{I}{s_{\alpha_0} -\sum_{i=1}^na_is_{\alpha_i}}&=\frac{s_{-\alpha_0}}{I -\sum_{i=1}^n a_i s_{-(\alpha_0-\alpha_i)}}\\
=&x^{\alpha_0-1}E_{(\alpha_0-\alpha_1, \cdots \alpha_0-\alpha_n), \alpha_0}(a_1 x^{\alpha_0-\alpha_1}, \cdots, a_n x^{\alpha_0-\alpha_n}).    \label{32}
\end{align}
Therefore we obtain
\begin{equation*}
y(x)=x^{\alpha_0-1}E_{(\alpha_0-\alpha_1, \cdots \alpha_0-\alpha_n), \alpha_0}(a_1 x^{\alpha_0-\alpha_1}, \cdots, a_n x^{\alpha_0-\alpha_n}) \star g(x).
\end{equation*}
Thus, we have the solution (29).
Now we consider  $y(x) \in \bigcap^n_{i=0} \Omega ^{\alpha_i}_{\beta_i}$.
Let $y_{\gamma_k}:=I^{\gamma_k}_{0+}y$ for $\gamma_k=(1-\beta_k)(m_k-\alpha_k),~k \in \mathbf{Z}_{0,n}$.
Then we have 
\begin{align*}
y_{\gamma_k}&=I^{\gamma_k}_{0+}((x^{\alpha_0-1}E_{(\alpha_0-\alpha_1, \cdots, \alpha_0-\alpha_n), \alpha_0}(a_1 x^{\alpha_0-\alpha_1}, \cdots, a_n x^{\alpha_0-\alpha_n})) \star g(x)) \\
=&h_{\gamma_k}(x) \star (x^{\alpha_0-1}E_{(\alpha_0-\alpha_1, \cdots, \alpha_0-\alpha_n), \alpha_0}(a_1 x^{\alpha_0-\alpha_1}, \cdots, a_n x^{\alpha_0-\alpha_n})) \star g(x) \\
=& (x^{\alpha_0-1}E_{(\alpha_0-\alpha_1, \cdots, \alpha_0-\alpha_n), \alpha_0}(a_1 x^{\alpha_0-\alpha_1}, \cdots, a_n x^{\alpha_0-\alpha_n})) \star (h_{\gamma_k}(x) \star g(x)) \\
=& \int_{0}^{x}t^{\alpha_0-1}E_{(\alpha_0-\alpha_1, \cdots, \alpha_0-\alpha_n), \alpha_0}(a_1 t^{\alpha_0-\alpha_1}, \cdots, a_n t^{\alpha_0-\alpha_n})g_{\gamma_k}(x-t)dt
\end{align*}
where 
\[g_{\gamma_k}(x):=I^{\gamma_k}_{0+}g(x) \in \left \{
\begin{array}{rl}
C_{-1} & \text{if}\quad \alpha_0 \in \mathbf{N}, \\
C^1_{-1} & \text{if}\quad \alpha_0 \notin \mathbf{N}. \\
\end{array} \right.  \]
Whereas, from (31) we have 
\begin{equation}
y(x)=I^{\gamma_0}_{0+}g(x) + \sum _{i=1}^{n}a_iI^{\gamma_0-\gamma_i}_{0+}y(x)    \label{33}
\end{equation}
and
\begin{equation}
y_{\gamma_k}(x)=I^{\gamma_0}_{0+}g_{\gamma_k}(x) + \sum _{i=1}^{n}a_iI^{\gamma_0-\gamma_i}_{0+}y_{\gamma_k}(x).    \label{34}
\end{equation}
By (34) we obtain 
\begin{equation}
y_{\gamma_k}(x)=(I^{\gamma_0\-\gamma_1}_{0+}\psi_1)(x),~\psi_1 \in \left\{
\begin{array}{rl}
C_{-1} & \text{if}\quad \alpha_0 \in \mathbf{N}, \\
C^1_{-1} & \text{if}\quad \alpha_0 \notin \mathbf{N}. \\
\end{array} \right.                    \label{35}
\end{equation}
Combining now the relations (34) and (35) and repeating the same arguments $p$ times $(p=[\alpha_0/(\alpha_0-\alpha_1)]+1)$, we arrive at the representation 
\begin{equation}
y_{\gamma_k}(x)=(I^{\gamma_0}_{0+}\psi_p)(x),~\psi_p \in \left\{
\begin{array}{rl}
C_{-1} & \text{if}\quad \alpha_0 \in \mathbf{N}, \\
C^1_{-1} & \text{if}\quad \alpha_0 \notin \mathbf{N}. \\
\end{array} \right.                   \label{36}
\end{equation}
In the case $\alpha_0=m_0 \in \mathbf{N}$, it follows from (36) that
\begin{equation*}
y^{(m_0)}_{\gamma_0}=\psi_p \in C_{-1},~y_{\gamma_0}(0)= \cdots =y^{(m_0-1)}_{\gamma_0}(0)=0.
\end{equation*}
If $\alpha_0 \notin \mathbf{N}$ then since
\begin{equation*}
h_{\gamma_0}(x)=\frac{x^{\gamma_0-1}}{\Gamma(\gamma_0)} \in C^{m_0-1}_{-1},~h_{\alpha_0}(0)= \cdots =h^{(m_0-2)}_{\alpha_0}(0)=0,
\end{equation*}
we obtain $y_{\gamma_0}\in C_{-1}^{m_0},~y_{\gamma_0}(0)=\cdots=y_{\gamma_0}^{(m_0-1)}(0)=0$ by using Lemma 2.
In the case $k=1,\cdots,n$  similarly with above argument, we have
\begin{equation*}
y_{\gamma_k} \in C^{m_k}_{-1},~y_{\gamma_k}(0)= \cdots =y^{(m_k-1)}_{\gamma_k}(0)=0.
\end{equation*}
Hence $y(x)$ is a unique solution of homogeneous initial value problem (25) and (28).
The proof of theorem is completed.(QED)
%%%%\end{pr}

%---------------------Theorem 2----------------------
\begin{thr}
For the initial value problem (25) and (26) with inhomogeneous initial conditions we have the following conclusions. 

i) If $(1-\beta_0)(m_0-\alpha_0)= \cdots =(1-\beta_n)(m_n-\alpha_n)=\gamma$, then the initial value solution(25) and (26) has the unique solution 
\begin{align}
\nonumber &y(x)=y_g(x)+\sum^{m_0-1}_{k=0}y_{k-\gamma}\left\{\frac{x^{k-\gamma}}{\Gamma(k-\gamma+1)}+\right. \\
&+\sum^{n}_{i=l_k+1}\left.a_ix^{k-\gamma+\alpha_0-\alpha_i}E_{(\alpha_0-\alpha_1, \cdots, \alpha_0-\alpha_n), k+1-\gamma+\alpha_0-\alpha_i}(a_1 x^{\alpha_0-\alpha_1}, \cdots, a_n x^{\alpha_0-\alpha_n})\right\}   \label{37}
\end{align}
in the space $\bigcap ^n_{i=0} \Omega ^{\alpha_i}_{\beta_i}$, where $l_k \in \mathbf{N}_0$ for every fixed $k=0, \cdots ,m_0-1$ is the integer such that
\begin{equation}
m_{l_k} \geq k+1\quad\text{and}\quad m_{l_k+1} \leq k.       \label{38}
\end{equation}
(If $m_i \leq k,~i=0,\cdots,n$ we set $l_k:=0$, and if $m_i \geq k+1,~i=0, \cdots ,n$, then we let $l_k:=n$.)

ii) If $\gamma=(1-\beta_0)(m_0-\alpha_0)= \cdots =(1-\beta_{l-1})(m_{l-1}-\alpha_{l-1}) \neq (1-\beta_l)(m_l-\alpha_l),~0<l \leq n$, then there exists a solution $y(x)$ in the space $\bigcap^n_{i=0} \Omega ^{\alpha_i}_{\beta_i}$ if and only if 
\begin{align}
\nonumber &y_{k-(1-\beta_i)(m_i-\alpha_i)}=0,~i \in \mathbf{Z}_{l,n},~k \in \mathbf{Z}_{0,~m_i-1},\\ 
\nonumber &y_{k-(1-\beta_i)(m_i-\alpha_i)}=0,~i \in \mathbf{Z}_{0,l-1},~k \in \mathbf{Z}_{0,~M}\\
&(where~M=[m_l-(1-\beta_l)(m_l-\alpha_l)-1+\gamma])          \label{39}
\end{align}
and a unique solution $y(x)$ is provided by
\begin{align}
\nonumber &y(x)=y_g(x)+\sum^{m_0-1}_{k=M}y_{k-\gamma}\left\{\frac{x^{k-\gamma}}{\Gamma(k-\gamma+1)}+\right.\\
&+\sum^{n}_{i=l_k+1}\left.a_ix^{k-\gamma+\alpha_0-\alpha_i}E_{(\alpha_0-\alpha_1, \cdots, \alpha_0-\alpha_n), k+1-\gamma+\alpha_0-\alpha_i}(a_1 x^{\alpha_0-\alpha_1}, \cdots, a_n x^{\alpha_0-\alpha_n})\right\},  \label{40}
\end{align}
where $l_k \in \mathbf{N}_0,~k=0,\cdots, m_0-1$ are determined from the condition
\begin{equation}
m_{l_k} \geq k+1 \quad and \quad m_{l_k+1} \leq k .       \label{41}
\end{equation}
(If $m_i \leq k,~i=0,\cdots,l-1$ we set $l_k:=0$, and if $m_i \geq k+1,~i=0,\cdots, l-1$, then we let $l_k:=l-1$. In (37) and (40) $y_g(k)$ is the solution (29) of homogeneous initial value problem (25) and (28).
\end{thr}
%%%%%%------------------proof----------------
%%%\begin{pr}
~\textbf{Proof:}~By using lemma 4 and lemma5 we get the following algebraic equation in the field $M_{-1}$ of convolution quotients
\begin{equation}
s_{\alpha_0} \cdot y-\sum_{i=1}^n a_i s_{\alpha_i} \cdot y=g+s_{\alpha_0} \cdot y_{\alpha_0,\beta_0}-\sum_{i=1}^n a_i s_{\alpha_i} \cdot y_{\alpha_i,\beta_i}  \label{42}
\end{equation}
from inhomogeneous initial value problem (25) and (26), where $y_{\alpha_i,\beta_i}$ is represented by (14).
This linear equation has a unique solution in the field $M_{-1}$ 
\begin{equation}
y=\frac{I}{s_{\alpha_0} -\sum_{i=1}^n a_i s_{\alpha_i}} \cdot g + \frac{I}{s_{\alpha_0} -\sum_{i=1}^n a_i s_{\alpha_i}}\left(s_{\alpha_0} \cdot y_{\alpha_0,\beta_0}-\sum_{i=1}^n a_i s_{\alpha_i} \cdot y_{\alpha_i,\beta_i}\right). \label{43}
\end{equation}
Here $y_g:=\frac{I}{s_{\alpha_0} -\sum_{i=1}^n a_i s_{\alpha_i}} \cdot g$ is represented by (29) and is the solution of homogeneous initial value problem (25) and (28). Let 
\begin{equation}
y_h:=\frac{I}{s_{\alpha_0} -\sum_{i=1}^n a_i s_{\alpha_i}} \cdot \left(s_{\alpha_0} \cdot y_{\alpha_0,\beta_0}-\sum_{i=1}^n a_i s_{\alpha_i} \cdot y_{\alpha_i,\beta_i}\right).\label{44}
\end{equation}
Then we have 
\begin{align*} 
y_h = \frac{I}{I -\sum_{i=1}^n a_i s_{\alpha_i-\alpha_0}}\left\{\sum^{m_0-1}_{k=0}y_{k-(1-\beta_0)(m_0-\alpha_0)}~s_{-(k-(1-\beta_0)(m_0-\alpha_0)+1)}-\right. 
\end{align*}
\begin{equation}
- \sum^n_{i=1}a_i s_{\alpha_i-\alpha_0}\left.\sum^{m_0-1}_{k=0}y_{k-(1-\beta_0)(m_0-\alpha_0)}~s_{-(k-(1-\beta_0)(m_0-\alpha_0)+1)}\right\}.\label{45}
\end{equation}
%%%%%\end{eqnarray}

If $(1-\beta_0)(m_0-\alpha_0)= \cdots =(1-\beta_n)(m_n-\alpha_n)=\gamma$, from (45) we have 
\begin{align*}
y_h&=\sum^{m_0-1}_{k=0}y_{k-\gamma} \frac{I}{s_{k-\gamma+1}} \cdot \frac{I-\sum^{l_k}_{i=1}a_i s_{\alpha_i-\alpha_0}}{I-\sum^{n}_{i=1} a_i s_{\alpha_i-\alpha_0}}\\
&=\sum^{m_0-1}_{k=0}y_{k-\gamma}\frac{I}{s_{k-\gamma+1}}\cdot\left\{I+\frac{\sum^{n}_{i=l_k+1} a_i s_{\alpha_i-\alpha_0}}{I-\sum^{n}_{i=1} a_i s_{\alpha_i-\alpha_0}}\right\}\\
&=\sum^{m_0-1}_{k=0}y_{k-\gamma}\left\{\frac{x^{k-\gamma}}{\Gamma(k-\gamma+1)}+\sum^n_{i=l_k+1} a_i x^{k-\gamma+\alpha_0-\alpha_i}E_{\cdot, k+1-\gamma+\alpha_0-\alpha_i}(x)\right\},
\end{align*}
where $l_k,~k=0, \cdots, m_0-1$ are determined from condition (38).
According to the definition of the numbers $l_k$, we have $m_i \leq k$ for $i \in \mathbf{Z}_{l_k+1,n}$. It follows then that $k-\gamma +\alpha_0-\alpha_i \geq \alpha_0-\gamma,~i \in \mathbf{Z}_{l_k+1,n}$.
Using this inequality, we readily get the inclusion $I^{\gamma}_{0+}y_h \in C^{m_0}_{-1}$ and $y(x) \in \bigcap ^n_{i=0} \Omega ^{\alpha_i}_{\beta_i}$.
For $u_k:=\frac{x^{k-\gamma}}{\Gamma(k-\gamma+1)}+\sum^n_{i=l_k+1} a_i x^{k-\gamma+\alpha_0-\alpha_i} E_{(\cdot), k+1-\gamma+\alpha_0-\alpha_i}(x)$ we have
\begin{equation*}
(I^{\gamma}_{0+}u_k)^{(l)}(0)= \left\{
\begin{array}{rl}
1 & \text{if}\quad k=l \\
0 & \text{if}\quad k \neq l \\
\end{array} \right.
, k,l=0, \cdots, m_0-1.
\end{equation*}
Therefore $y_h$ satisfies the initial condition (26) and hence $y(x)=y_g(x)+y_h(x)$ is a unique solution of the initial value problem (25) and (26).

If $\gamma=(1-\beta_0)(m_0-\alpha_0)= \cdots =(1-\beta_{l-1})(m_{l-1}-\alpha_{l-1}) \neq (1-\beta_l) (m_l-\alpha_l),~0<l \leq n$, then from (45) we have 
\begin{align*}
y_h=\frac{I}{I-\sum^{n}_{i=1} a_i s_{\alpha_i-\alpha_0}} \cdot &\left\{\sum^{m_0-1}_{k=0}\right.y_{k-\gamma}~s_{-(k-\gamma+1)}\\
&\quad\quad\quad\quad\quad-\sum^{l-1}_{i=1}a_i s_{\alpha_i-\alpha_0} \sum^{m_i-1}_{k=0}y_{k-\gamma}~s_{-(k-\gamma+1)}-
\end{align*}
\begin{equation*}
-\left.\sum^n_{i=l} a_i s_{\alpha_i-\alpha_0}\sum^{m_i-1}_{k=0}y_{k-(1-\beta_i)(m_i-\alpha_i)}~s_{-(k-(1-\beta_i)(m_i-\alpha_i)+1)}\right\}
\end{equation*}
\begin{align*}
&=\sum^{m_0-1}_{k=0} y_{k-\gamma} s_{-(k-\gamma+1)} \frac{I-\sum^{l_k}_{i=1} a_i s_{\alpha_i-\alpha_0}} {I-\sum^{n}_{i=1} a_i s_{\alpha_i-\alpha_0}}  \\
&-\frac{I}{I-\sum^{n}_{i=1} a_i s_{\alpha_i-\alpha_0}} \sum^{n}_{i=l}a_i s_{\alpha_i-\alpha_0} \sum^{m_i-1}_{k=0}y_{k-(1-\beta_i)(m_i-\alpha_i)}s_{-(k-(1-\beta_i)(m_i-\alpha_i)+1)} 
\end{align*}
where $l_k$ is defined by (41).  From the above equation we have 
\begin{align*}%%%%\begin{equation*}
y_h&(x)=\sum^{m_0-1}_{k=0} y_{k-\gamma}\left\{\frac{x^{k-\gamma}}{\Gamma(k-\gamma+1)}+\sum^n_{i=l_k+1} a_i x^{k-\gamma+\alpha_0-\alpha_i}E_{(\cdot), k+1-\gamma+\alpha_0-\alpha_i}(x)\right\}\\
-& \sum^{n}_{i=l} a_i \sum^{m_i-1}_{k=0} y_{k-(1-\beta_i)(m_i-\alpha_i)}x^{\alpha_0-\alpha_i+k-(1-\beta_i)(m_i-\alpha_i)}E_{(\cdot), \alpha_0-\alpha_i+k+1-(1-\beta_i)(m_i-\alpha_i)}(x). 
\end{align*}    %\end{equation*}
In order that $y_h(x) \in \bigcap ^n_{i=0} \Omega ^{\alpha_i}_{\beta_i}$ , the following relation should be satisfies 
\begin{align}
\alpha_0-\alpha_i+k+1-(1-\beta_i)(m_i-\alpha_i)+\gamma-m_0=\nonumber \\
\quad =k+1-m_i+\beta_i(m_i-\alpha_i)-\beta_0(m_0-\alpha_0) \geq 0\nonumber
\end{align}
for every $i \in \mathbf{Z}_{l,n},~k \in \mathbf{Z}_{0,m_i-1}$. This relation satisfies only in the case 
\[k=m_i-1~\text{and}~\beta_i(m_i-\alpha_i) \geq \beta_0(m_0-\alpha_0).\]
Therefore $y_{k-(1-\beta_i)(m_i-\alpha_i)}=0$ for every $i \in \mathbf{Z}_{l,n},~k \in \mathbf{Z}_{0,m_i-2}$ and, as $y_h(x)$ must satisfies the initial condition (26), we get $y_{m_i-1-(1-\beta_i)(m_i-\alpha_i)}=0$ for every $i \in \mathbf{Z}_{l,n}$.
Also, as the relation 
\[k-\gamma+(1-\beta_l)(m_l-\alpha_l)-k_1 \geq 0, ~k=0, \cdots, m_0-1,~k_1=0, \cdots, m_l-1\]
must satisfies , the following relations should be satisfies 
\[y_{k-\gamma}=0,~ k=0,\cdots,M,~\text{where}~M=[m_l-(1-\beta_l)(m_l-\alpha_1)-1+\gamma].\]
Therefore we obtain 
\begin{align*}
y_h(x)=\sum^{m_0-1}_{k=M} y_{k-\gamma}&\left(\frac{x^{k-\gamma}}{\Gamma(k-\gamma+1)}+\right.\\
&+\sum^n_{i=l_k+1} \left.a_i x^{k-\gamma+\alpha_0-\alpha_i}E_{(\cdot), k+1-\gamma+\alpha_0-\alpha_i}(x)\right).
\end{align*}
Hence the solution is $y(x)=y_g(x)+y_h(x)$.
The proof of theorem 2 is completed.(QED)

%%%%\end{pr}
%%-----------------------Example 1-------------------------
\begin{exam}
If $\beta_0= \cdots = \beta_n=1$ then the initial value problem (25) and (26) becomes the initial value problem with Caputo fractional derivatives
\begin{equation}
^cD^{\alpha_0}_{0+}y(x)-\sum^n_{i=1}a_i ^cD^{\alpha_i}_{0+}y(x)=g(x), \label{46}
\end{equation}
\begin{equation}
\frac{d^k}{dx^k}y(0+)=y_{k} \in \mathbf{R},~ k \in \mathbf{Z}_{0,~m_0-1},\label{47}
\end{equation}
where $\alpha_0>\alpha_1> \cdots >\alpha_n \geq 0,~m_i-1<\alpha_i \leq m_i \in \mathbf{N},~i \in \mathbf{Z}_{0,n}$, and the space $\bigcap ^n_{i=0} \Omega ^{\alpha_i}_{\beta_i}$ is coincides with $C^{m_0}_{-1}$. From the theorem 2 the initial value problem (46) and (47) has a unique solution 
\begin{align}    %\begin{equation}
y(&x)=\int^{x}_{0}t^{\alpha_0-1}E_{(\alpha_0-\alpha_1, \cdots,\alpha_0-\alpha_n), \alpha_0}(a_1 t^{\alpha_0-\alpha_1}, \cdots, a_n t^{\alpha_0-\alpha_n})g(x-t)dt\nonumber \\
+&\sum^{m_0-1}_{k=0}y_{k}\left[\frac{x^{k}}{\Gamma(k+1)}+\right. \nonumber \\
&+\sum^{n}_{i=l_k+1}\left. a_i x^{k+\alpha_0-\alpha_i}E_{(\alpha_0-\alpha_1, \cdots, \alpha_0-\alpha_n), k+1+\alpha_0-\alpha_i}(a_1 x^{\alpha_0-\alpha_1}, \cdots, a_n x^{\alpha_0-\alpha_n})\right], \label{48}    
\end{align}    %\end{equation}
where $l_k,~k=0, \cdots, m_0-1$ are determined from condition (38).
\end{exam}
%----------------Example 2-------------------
\begin{exam}
If $\beta_0= \cdots =\beta_n=0$, then the initial value problem (25) and (26) becomes the initial value problem with Riemann-Liouville fractional derivatives
\begin{align}
&D^{\alpha_0}_{0+}y(x)-\sum^n_{i=1}a_i D^{\alpha_i}_{0+}y(x)=g(x), \label{49}\\
&\frac{d^k}{dx^k} I^{m_i-\alpha_i}_{0+}y(0+)=y_{k-(m_i-\alpha_i)} \in \mathbf{R},~ i \in \mathbf{Z}_{0,n},~ k \in \mathbf{Z}_{0,m_i-1},        \label{50}
\end{align}
where $\alpha_0>\alpha_1> \cdots >\alpha_n \geq 0,~m_i-1<\alpha_i \leq m_i \in \mathbf{N},~i \in \mathbf{Z}_{0,n}$ .

If $m_0-\alpha_0= \cdots =m_n-\alpha_n=\gamma$ then the initial value problem (49) and (50) has unique solution 
\begin{align}    %\begin{equation}
y(x)=&\int^{x}_{0}t^{\alpha_0-1}E_{(\alpha_0-\alpha_1, \cdots,\alpha_0-\alpha_n), \alpha_0}(a_1 t^{\alpha_0-\alpha_1}, \cdots, a_n t^{\alpha_0-\alpha_n})g(x-t)dt +\nonumber \\
+&\sum^{m_0-1}_{k=0}y_{k-\gamma}\left[\frac{x^{k-\gamma}}{\Gamma(k-\gamma+1)}+\sum^{n}_{i=l_k+1} a_i x^{k-\gamma+\alpha_0-\alpha_i}E(\cdots) \right].                              \label{51}
\end{align}    %\end {equation}
$($Here $E(\cdots)$ = $E_{(\alpha_0-\alpha_1, \cdots \alpha_0-\alpha_n),~ k+1-\gamma+\alpha_0-\alpha_i} 
(a_1 x^{\alpha_0-\alpha_1}, \cdots, a_n x^{\alpha_0-\alpha_n}).)$ in the space $\bigcap ^n_{i=0} \Omega ^{\alpha_i}_{0}=\Omega ^{\alpha_i}_{0}$. Here $l_k,~k=0, \cdots, m_0-1$ are determined from condition (41).

If $\gamma=m_0-\alpha_0= \cdots =m_{l-1}-\alpha_{l-1} \neq m_l- \alpha_l,~0<l \leq n$ then exists a solution $y(x)$ in the space $\bigcap ^n_{i=0} \Omega ^{\alpha_i}_{0}=\Omega ^{\alpha_i}_{0}$, if and only if 
\begin{align}    %\begin{equation}
&y_{k-(m_i-\alpha_i)}=0,~i \in \mathbf{Z}_{l,n},~k \in \mathbf{Z}_{0,m_i-1}  \nonumber \\
&y_{k-(m_i-\alpha_i)}=0,~i \in \mathbf{Z}_{0,l-1},~k \in \mathbf{Z}_{0,M},~M=[\alpha_l-1+\gamma].    \label{52}
\end{align}    %\end{equation}
Then the unique solution of the initial value problem (49) and (50) is provided by
\begin{align}    %\begin{equation}
y(x)=&\int^{x}_{0}t^{\alpha_0-1}E_{(\alpha_0-\alpha_1, \cdots \alpha_0-\alpha_n), \alpha_0}(a_1 t^{\alpha_0-\alpha_1}, \cdots, a_n t^{\alpha_0-\alpha_n})g(x-t)dt+\nonumber \\
+&\sum^{m_0-1}_{k=M}y_{k-\gamma}\left[\frac{x^{k-\gamma}}{\Gamma(k-\gamma+1)}+\sum^{n}_{i=l_k+1} a_i x^{k-\gamma+\alpha_0-\alpha_i}E(\cdots)\right].    \label{53}
\end{align}    %\end{equation}
Here $E(\cdots)$ and $l_k,~k=0, \cdots ,m_0-1$ are the same as in (51).
\end{exam}
%---------------------------Example 3---------------
\begin{exam}
We consider the so-called composite fractional relaxation equation (see \cite{hil3})
\begin{align}
&\tau_1 \frac{d}{dt}f(t) + \tau^{\alpha}_2 D^{\alpha,\mu}_{0+}f(t) + f(t)=0,    \label{54}\\
&f(0+)=1 ,~(I^{(1-\mu)(1-\alpha)}_{0+}f)(0+)=f_{(1-\mu)(1-\alpha)},    \label{55}
\end{align}
where $0<\alpha<1,~0 \leq \mu \leq 1,~0<\tau_1,~< \infty$.  
By using the theorem 2 the initial value problem (54) and (55) have a unique solution $f(t)$ in the space $\Omega^1_0 \bigcap \Omega ^{\alpha}_{\mu} \bigcap \Omega ^0_0$, if and only if 
\begin{equation}
(I^{(1-\mu)(1-\alpha)}_{0+}f)(0+)=f_{(1-\mu)(1-\alpha)}=0.    \label{56}
\end{equation}
Then the unique solution is given by
\begin{equation}
f(t)=E_{(1-\alpha,~1),~1}(-\frac{\tau^{\alpha}_2}{\tau_1} t^{1-\alpha},~-\frac{1}{\tau-1}t),~t>0.    \label{57}
\end{equation}
\end{exam}

{\bf Acknowledgment} The authors would like to thank Prof. Virginia Kiryakova and Prof. Dr. Yuri Luchko for their help and advices for the improvement of this article.

%----------------------References

\end{document}